\documentclass{aa} % for a referee version
\usepackage{graphicx}
\usepackage{amssymb}
\usepackage{amsmath}
\begin{document}
\newcommand{\kmpsb}{km~s$^{-1}$ }
\newcommand{\cdhob}{C$^{18}$O }
\newcommand{\cdsob}{C$^{17}$O }
\newcommand{\dzcob}{$^{12}$CO }
\newcommand{\tzcob}{$^{13}$CO }
\newcommand{\ndhpb}{N$_2$H$^+$ }
\newcommand{\cdb}{column density }
\newcommand{\ccb}{cm$^{-3}$ }
\newcommand{\cdeb}{cm$^{-2}$ }
\newcommand{\scb}{cm$^{-2}$ }
\newcommand{\ctdsb}{C$^{32}$S }
\newcommand{\ctqsb}{C$^{34}$S }
\newcommand{\tdsob}{$^{32}$SO }
\newcommand{\tqsob}{$^{34}$SO }
\newcommand{\juzb}{(J:1--0) }
\newcommand{\jdub}{(J:2--1) }
\newcommand{\jtdb}{(J:3--2) }
\newcommand{\denb}{n(H$_2$) }
\newcommand{\kmps}{km~s$^{-1}$}
\newcommand{\cdho}{C$^{18}$O}
\newcommand{\cdso}{C$^{17}$O}
\newcommand{\dzco}{$^{12}$CO}
\newcommand{\tzco}{$^{13}$CO}
\newcommand{\ndhp}{N$_2$H$^+$}
\newcommand{\nddp}{N$_2$D$^+$}
\newcommand{\nddpb}{N$_2$D$^+$ }
\newcommand{\cd}{column density}
\newcommand{\cc}{cm$^{-3}$}
\newcommand{\cde}{cm$^{-2}$}
\newcommand{\ctds}{C$^{32}$S}
\newcommand{\ctqs}{C$^{34}$S}
\newcommand{\tdso}{$^{32}$SO}
\newcommand{\tqso}{$^{34}$SO}
\newcommand{\juz}{(J:1--0)}
\newcommand{\jdu}{(J:2--1)}
\newcommand{\jtd}{(J:3--2)}
\newcommand{\den}{n(H$_2$)}
\newcommand{\mjy}{MJy/sr}%$^{-1}$}
\newcommand{\mjyb}{MJy/sr }%$^{-1}$ }
\newcommand{\Av}{A$_{\mathrm V}$}
\newcommand{\Avb}{A$_{\mathrm V}$ }
\newcommand{\SM}{M$_\odot$}
\newcommand{\SMb}{M$_\odot$ }
   \title{Comments on the paper ``The initial conditions of isolated
   star formation - VI. SCUBA mapping of prestellar cores'' (Kirk et
   al. \cite{Kirk05})} \author{L. Pagani \inst{1}, G. Lagache \inst{2}}

\offprints{L. Pagani}

   \institute{LERMA \& UMR 8112 du CNRS, Observatoire de Paris, 61, Av. de l'Observatoire
   F-75014 Paris\and IAS,
   B{\^{a}}t. 121, Universit\'e Paris-Sud F-91435 Orsay\\ \email{laurent.pagani@obspm.fr}}
   \date{\today}
\titlerunning{Comments on Kirk et al. (\cite{Kirk05})}
\authorrunning{Pagani L.}
   \abstract{In their survey paper of prestellar cores with SCUBA,
   Kirk et al. (\cite{Kirk05}) have discarded two of our papers on
   L183 (Pagani et al. \cite{Pagani03}, \cite{Pagani04}). However
   these papers bring two important pieces of information that they
   cannot ignore. Namely, the real structure of L183 and the very poor
   correlation between submillimeter and far infrared (FIR) dust
   emission beyond \Avb $\approx$ 15 mag. Making the erroneous
   assumption that it is the same dust that we are seeing in emission 
   at both 200 and 850 $\mu$m, they derive constant temperatures
   which are only approximate, and column densities which are too
   low. In fact dust temperatures do decrease inside dark clouds and
   the FIR emission is only tracing the outer parts of the dark
   clouds (Pagani et al. \cite{Pagani04}).}
\maketitle
   \keywords{ISM: dust,extinction
-- ISM: Structure -- ISM: individual: L134N -- ISM: individual: L183
}
\section{The L183 case}
\subsection{Historical background}
Ward-Thompson et al. (\cite{Ward94}) reported a first submm source in
L183 centered at 15h54m -2\degr51\arcmin\ (J2000) and Ward-Thompson et
al. (\cite{Ward00}) reported a second submm source centered at
15h54m09 -2\degr52\arcmin38\arcsec\ (J2000), about 90\arcsec\ further
south. The existence of the first source is not mentioned in the
second paper. Lehtinen et al. (\cite{Lehtinen03}) combined these two
detections as two separated sources which they identified with FIR
peaks from an ISOPHOT 200 $\mu$m strip despite a difference of 30\arcsec\
in the separation of the two sources between the submm and the FIR
identifications. 

We then showed (Pagani et al. \cite{Pagani03}) from a large MAMBO map
and from an ISOCAM absorption map that there was no northern submm
source compatible with the position reported in Ward-Thompson et
al. (\cite{Ward94}). This was also confirmed by recent SCUBA maps at
850 $\mu$m from our own work (unpublished) and from Kirk et
al. (\cite{Kirk05}) present work.

We also showed that the 200 $\mu$m sources found by Lehtinen et
al. (\cite{Lehtinen03}) were in fact artefacts from their data reduction
process and that no point source could be clearly identified (Pagani
et al. \cite{Pagani03}, \cite{Pagani04}). Because of these
mis-identifications, most or all of the subsequent results discussed
in that paper are not valid.

\subsection{Present paper (Kirk et al. \cite{Kirk05})}
Kirk et al. (\cite{Kirk05}) report one point source at 850 $\mu$m and
two point sources at 450 $\mu$m, one in common with the 850 $\mu$m
peak and which coincides also with our own ISOCAM and MAMBO detection
and a second, further north for which they have no explanation apart
from a possible temperature gradient related to the fact that the 200
$\mu$m peak is situated even further north than this second peak.
Though this second peak could be approximately consistent with their first
detection (Ward-Thompson et al. \cite{Ward94}), or with Lehtinen et
al. (\cite{Lehtinen03}) source FIR1, they do not discuss the validity of their
detection with respect to these works. However, in their introduction,
they indicate that the results found by Lehtinen et al. are consistent
with their own findings.

Can we check the possibility of a source appearing at 450 $\mu$m (with
a 10 $\sigma$ detection) with no or at most weak counterpart at 850
$\mu$m ? First, let us evaluate the 450/850 $\mu$m ratio : the highest
contour is 1170 mJy/beam at 450 $\mu$m and 120 mJy/beam at 850 $\mu$m,
these contours have approximately the same size except that the 850
$\mu$m contour is obviously not a closed contour around this putative
source. If we suppose that the dust emissivity law varies with
$\lambda^{-2}$, we find that this ratio of $\approx$ 10 is indicative
of a 35 K source. This is a large value for a dark cloud which could
indicate the presence of an embedded protostar. However this hot spot
is not seen with IRAS at 100 $\mu$m nor with ISOPHOT at the same
wavelength despite its higher resolution (45\arcsec). Keeping the same
dust dependency with wavelength, the flux at 100 $\mu$m should be
$\approx$ 10$^4$ MJy sr$^{-1}$ if we extrapolate from the 450 $\mu$m
estimate. If we take a more pessimistic dust emissivity law
proportional to $\lambda^{-1.5}$ or varying from $\lambda^{-2}$ to
$\lambda^{-1}$ between 450 and 100 $\mu$m, this would lower the
flux only by a factor of 2. We must take into account the dilution of
this hot spot in the ISOPHOT beam. The hot spot is 15\arcsec\ wide
(i.e. the SCUBA resolution. The contours are slightly more extended in
fact but let us take this value as a lower limit) and the ISOPHOT
resolution is 45\arcsec\ thus even if the remaining dust in the
ISOPHOT pixel has a negligible emission flux, the hot spot emission
should still be 10$^3$ MJy sr$^{-1}$ at 100 $\mu$m and still 500 MJy
sr$^{-1}$ if we lower the emissivity of the dust in the FIR. It is
obvious that such a source could not be missed in the ISOPHOT 100
$\mu$m map which yields 27 MJy sr$^{-1}$ including the diffuse dust
emission and only 5 MJy sr$^{-1}$ after the diffuse emission has been
subtracted. One should remember that SCUBA is totally insensitive to
diffuse extended emission and there should be no reason to consider it
here. Even if we lower the dust temperature to 20 K instead of 35 K,
we find a final flux (including dilution) of 120 MJy sr$^{-1}$ (or 60
for a lower emissivity), several times more than the measured
value. Thus the northern 450 $\mu$m source is most probably an
artefact and the north-south temperature gradient is not a correct
explanation for this detection.
\section{The dust temperature constancy}
\begin{figure}[t!]
   \flushleft
   \includegraphics[width=6.0cm,angle=-90]{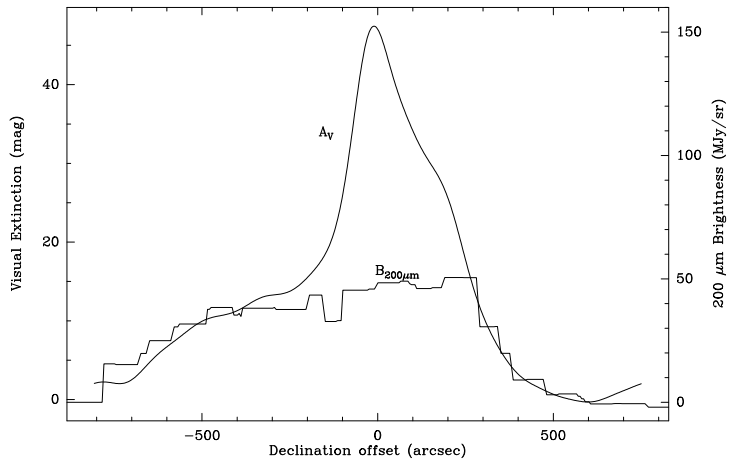}
   \caption{ISOPHOT 200 $\mu$m emission (labeled B$_{200\mu}$) seen
   along a north--south cut through the main dust peak and compared to
   the dust measured in absorption (labeled A$_\mathrm{V}$) and
   degraded to 90\arcsec\ resolution. The 200 $\mu$m emission
   perfectly matches the extinction due to the dust on the outer parts
   of the cloud but quickly flattens out in the middle. The loss of
   correlation between the dust in absorption and in emission is
   striking (from Pagani et al. \cite{Pagani04}) }
\label{cut}%
\end{figure}

We have shown in both our papers, (Pagani et al. \cite{Pagani03},
\cite{Pagani04}) that the 200 $\mu$m was not tracing the coldest dust
but only its outskirts. The dust peaks are not seen and this is not a
question of beam dilution with ISOPHOT as we have shown specifically
in Pagani et al. (\cite{Pagani04}) by smoothing our dust map obtained
from near- and mid- infrared measurements to ISOPHOT resolution and
comparing it to the 200 $\mu$m map (see our Figs 5, 6 and 7 in
Pagani et al. \cite{Pagani04}). A north-south cut of both maps
(NIR+MIR map and ISOPHOT FIR map) to 
follow the profiles makes the case very clear. We reproduce here the
Fig. 6 of Pagani et al. (\cite{Pagani04}) to show this cut
(Fig. \ref{cut}). The most plausible explanation for this discrepancy
is that the temperature drops inside the cloud below 10 K. This is
predicted by Zucconi et al. (\cite{Zucconi}), Evans et
al. (\cite{Evans}) and Stamatellos \& Whitworth (\cite{Stamatellos})
and is contradictory with the assertion that constant temperature is
consistent with the observations. Our result is directly obtained from
the observations and does not rely on any of these models until we
want to quantify more precisely this effect. The criticism addressed
by Kirk et al. about the Zucconi model being inconsistent with the
observations is only partially right : deep inside the cores, the
temperature does not vary very much in any of these models, and the
variation is slow enough (or even constant, albeit at a level of about
7--7.5 K, after having dropped from 13--15 K outside the cloud,
following Stamatellos \& Whitworth \cite{Stamatellos}) to let
absorption profiles measured in the NIR and the emission profile in
the submm look similar. From our observations, it is clear that the
200 $\mu$m is tracing dust up to \Avb $\approx$ 15 mag (total along
the line of sight) and that beyond, the dust temperature has dropped
to such low temperatures that its contribution to the total 200 $\mu$m
emission is negligible. Thus the dust traced by submm and FIR emission
are not the same and deriving temperatures and subsequently dust
column densities from spectral energy distributions including both
sets of wavelengths is wrong. For example, in L183, we find a dust
temperature varying from $\approx$ 13 K to $\approx$ 7 K and a peak
opacity of 150 mag, instead of 10 K and 85 mag for Kirk et
al. (\cite{Kirk05}). Though the authors could argue that we are more
or less within the error bars they mention, we are at the edge of
these bars and the error is probably systematic in an overestimate of
the dust temperature and an underestimate of the dust column density
and mass for most of their sources for which they have combined submm
and FIR data.

\section{Conclusions}
Probably the main results of the paper of Kirk et al. (\cite{Kirk05})
are not much changed by these corrections we bring here and thus there
is all the less no benefit to discard works which partly contradict one's
results. At least our papers should have been discussed to say why
they disagree with our results, we are open to discussion. This would
help everybody to test and improve his/her arguments. There is also no
shame to recognize one's mistakes especially in the very difficult
domain of submm continuum observations and if the authors would do so
themselves it would avoid other people to use wrong data which always
adds noise to the debate and induce extra effort from other people to
correct and fight against the propagating mistakes. 

\end{document}